\documentclass[a4paper,11pt]{article}
\usepackage[dvips]{graphicx}
\usepackage{amssymb}
\usepackage{amsmath}

\begin{document}

\title{Corrections to Gravity due to a Sol Manifold Extra Dimensional Space}
\author{V.K.Oikonomou\thanks{
voiko@physics.auth.gr}\\
Dept. of Theoretical Physics Aristotle University of Thessaloniki,\\
Thessaloniki 541 24 Greece}
\maketitle

\begin{abstract}
The corrections to the gravitational potential due to a Sol extra
dimensional compact manifold, denoted as $M_A^3$, are studied. The
total spacetime is of the form $M^4\times M_A^3$. The range of the
Sol corrections is investigated and compared to the range of the
$T^3$ corrections. We find that for small values of the radius of
the extra dimensions ($R<10^{-6}$) the range of the Sol manifold
corrections are large compared to the corresponding 3-torus
corrections. Also, due to the rich parameter space, Sol manifolds
corrections can be larger, comparable or smaller in comparison to
the 3-torus case, for larger $R$.
\end{abstract}

\bigskip
\bigskip
\section*{Introduction}
It has been nearly ten years since the originating papers studying
$TeV$ scale extra dimensions \cite{arkani} appeared. After that,
the phenomenological implications of extra dimensional models were
extensively studied. Many of these studies are concentrated on the
modifications caused on the gravitational potential due to extra
dimensions \cite{kehagias,leontaris}. In general the modifications
are of Yukawa type (except in Randall-Sundrum warped extra
dimensional models), differing only in their strength and range.
In this letter we shall investigate the modification of the
gravitational potential caused by a Sol extra dimensional
manifold. Sol manifolds are compact and orientable three
dimensional manifolds, denoted as $M_A^3$. In the following
sections we shall exploit the Sol manifolds geometric structure
and the study of the Laplace equation on such manifolds. After
that we compare the range of the corrections to the gravitational
potential due to Sol manifolds, with the corresponding range of
the corrections caused by a $T^3$ extra dimensional manifold.

\section*{Sol Manifold Geometry and Topology}
Sol manifolds are one of Thurston's 8 three dimensional geometries
\cite{Thurston} (geometric structures). The geometrization
conjecture is an approach to the geometry of three dimensional
manifolds through general topological arguments. Actually one
starts with the fact that three dimensional manifolds compose from
2-spheres or torii surfaces. According to the properties and the
details of the topological gluing map of the above with $R$ or
$S^1$, the resulting manifolds acquire homogeneous metrics (at
least locally). Sol manifolds are described by one type of the 8
different homogeneous metrics.

Sol manifolds result from $SL(2,Z)$ stiffennings of torus bundles
over the circle. More elaborately a theorem \cite{tanimoto}
states: given M a total bundle space of a $T^2$ bundle over $S^1$
with gluing map $\phi$ and let $A$ $\epsilon$ $GL(2,Z)$ represent
the automorphism of the fundamental group of the torus $T^2$
induced by the gluing map $\phi$. Then the total bundle space
admits a Sol geometric structure if $A$ is hyperbolic, an $E^3$
structure if $A$ is periodic and finally a Nil structure
otherwise. $A$ is hyperbolic and gives rise to an orientable
manifold if $|TrA|>2$ and we shall dwell on this choice. More
formally, the Sol structure arises from the $SL(2,Z)$ stiffening
of the mapping torus of a torus diffeomorphism $\phi$
\cite{Thurston}.

Before proceeding let us make a comment on the structure of $A$.
Two different hyperbolic gluings with $A_1\neq A_2$, give rise to
different geometric Sol structures. For $A$ we will use the matrix
representation of the form
\begin{equation}\label{ASTRU}
A=\left(%
\begin{array}{cc}
  0 & 1 \\
  -1 & n \\
\end{array}%
\right)
\end{equation}
so the different classes of Sol geometric structures are
classified by the integer $n$, with $|n|=|TrA|>2$. However this
classification is not the only one existing \cite{kodama},
nevertheless this will do the classification properly in our case.
Also we shall take into account only the positive eigenvalues of
$A$.

Let us be more quantitative on the Sol manifold analysis. Consider
the manifold $T^2\times \mathbb{R}$ described by the periodic
coordinates $(x,y)$ for the torus, defined modulo $R$ (the radius
of the compact dimensions) and $z$ $\epsilon$ $(-\infty,\infty)$
be a coordinate of $\mathbb{R}$. The combined action of the torus
mapping through the hyperbolic gluing map $A$ diffeomorphism
(which we denote as $\bar{\Gamma}$) is :
\begin{equation}\label{mapptorus}
\bar{\Gamma}:\left(
\begin{array}{c}
  x \\
  y \\
  z \\
\end{array}
\right)
\rightarrow \left(%
\begin{array}{c}
  a_{11}x+a_{12}y \\
  a_{21}x+a_{22}y \\
  z+2\pi R \\
\end{array}%
\right)
\end{equation}
We give a general notation for the matrix $A$, of the form
\begin{equation}\label{generalm}
    A=\left(%
\begin{array}{cc}
  a_{11} & a_{12} \\
  a_{21} & a_{22} \\
\end{array}%
\right)
\end{equation}
but in practise we shall use the form of relation (\ref{ASTRU}).
As we described above, the Sol manifold, which is denoted as
$M_A^3$ is the quotient of $T^2\times \mathbb{R}$ by the action of
$\bar{\Gamma}$, that is $M_A^3\equiv T^2\times
\mathbb{R}/\bar{\Gamma}$. It is obvious that it is a total torus
bundle over $S^1$, with $T^2$ the fiber, $S^1$ the base space and
$A$ the hyperbolic gluing map of the torus fibers. We shall
describe the Sol manifolds for which the eigenvalues $\lambda$ of
$A$ are positive. For the case of (\ref{ASTRU}) the characteristic
polynomial of the matrix $A$ reads:
\begin{equation}\label{charpoly}
\lambda^2-TrA\lambda +1=0
\end{equation}
or equivalently:
\begin{equation}\label{charpol1}
\lambda^2-n\lambda +1=0
\end{equation}
The solutions to equation (\ref{charpol1}) are $\lambda$ and
$\lambda^{-1}$ with $\lambda+\lambda^{-1}=n=TrA$. Also the
discriminant of $A$ is:
\begin{equation}\label{discriminant}
D=(a_{11}+a_{22})^2-4
\end{equation}
which in our case reads:
\begin{equation}\label{discri1}
D=(\lambda -\lambda^{-1})^2
\end{equation}
In this letter we shall make use of another coordinate system
$(u,v,z)$ on the Sol manifold. The $(u,v)$ are linear coordinates
of the torus fibres related to an eigenbasis of the hyperbolic map
$A$. These coordinates correspond to a rotated torus lattice. The
action of $\bar{\Gamma}$ in the new coordinate system now reads:
\begin{equation}\label{newcoor}
\bar{\Gamma}:\left(
\begin{array}{c}
  u \\
  v \\
  z \\
\end{array}
\right)
\rightarrow \left(%
\begin{array}{c}
  \lambda u \\
  \lambda^{-1}v \\
  z+2\pi R \\
\end{array}%
\right)
\end{equation}
The original lattice was orthogonal while the new torus lattice is
not. If $e_u$ and $e_v$ are the basis of the lattice after the
action of $\bar{\Gamma}$ identification, then
$(e_u,e_v)=|e_u||e_v|\cos \theta$ (the basis of the new lattice
coincide with the eigenvectors of $A$ corresponding to the
eigenvalues $\lambda$ and $\lambda^{-1}$). Also the coordinates of
the fibers $T^2$ are not periodic anymore. So in order two pairs
$(u_1,v_1)$ and $(u_2,v_2)$ define the same point on the torus
lattice, the following must hold:
\begin{equation}\label{periodi}
(u_1-u_2,v_1-v_2)=ke_u+me_v
\end{equation}
with $k,m$ integers and $e_u$, $e_v$ defined previously.

\subsubsection*{Riemannian Sol group invariant metric on Sol
manifolds}

The Riemannian metric on Sol manifolds comes from the metric on
the universal covering of $M_A^3$. The group invariant metric on
the universal covering is a Sol group invariant metric. The Sol
group is a three dimensional solvable Lie group homeomorphic to
$\mathbb{R}^3$ and can be realized as the matrix:
\begin{equation}\label{solgroup}
\left(%
\begin{array}{ccc}
  e^{z\ln\lambda} & 0 & u \\
  0 & e^{-z\ln\lambda} & v \\
  0 & 0 & 1 \\
\end{array}%
\right)
\end{equation}
Finding the invariant metric on the universal covering of $M_A^3$
then using the relaxation method \cite{tanimoto} we can find the
following class of metrics of Sol manifolds (invariant under the
group (\ref{solgroup})):
\begin{equation}\label{metricsol}
\mathrm{d}s^2=Ee^{2z\ln\lambda}\mathrm{d}u^2+2F\mathrm{d}u\mathrm{d}v+Ge^{-2z\ln\lambda}\mathrm{d}v^2+\mathrm{d}z^2
\end{equation}
In the following we shall use the metric (\ref{metricsol}) in
order to compute the Laplace-Beltrami operator on $M_A^3$.

\noindent Sol manifolds are hyperbolic manifolds with negative
curvature. The only component of the Ricci tensor is $R_{zz}=-2$
for the metric (\ref{metricsol}), transformed in the coordinates
$x,y,z$. The hyperbolicity is a very interesting feature of Sol
manifolds, regarding they are torus fibrations. Due to their
hyperbolicity, the distribution of eigenvalues of the Laplacian is
not Poisson. This might have effects on the gravitons mass
splitting, as we discuss later.

\section*{Newton's law and extra dimensions}
The main purpose of this paper is to examine the corrections to
the gravitational potential caused by an extra dimensional Sol
manifold. The spacetime manifold is of the form $M^4\times M_A^3$.
In this section we review the general technique to obtain these
corrections. The presentation is based on reference
\cite{kehagias}. The techniques that will be presented are applied
to compact manifolds, since this is the case of $M_A^3$. A
generalization to non-compact manifolds can be found in
\cite{kahagias2}.

Consider a spacetime of the form $M^4\times M^n$, with $M^n$ an
n-dimensional compact manifold and $M^4$ the four dimensional
Minkowski spacetime. Suppose there exist a complete set of
orthogonal harmonic functions on $M^4$, $\Psi_m$, satisfying the
orthogonality condition:
\begin{equation}\label{orthogo}
\int_{M^n}\Psi_n(x)\Psi_m^*(x)=\delta_{n,m}
\end{equation}
and the completeness relation:
\begin{equation}\label{complet}
\sum_m\Psi_m(x)\Psi_m^*(x')=\delta^{(n)}(x-x')
\end{equation}
The functions $\Psi_m$ are eigenfunctions of the $n$-dimensional
Laplace-Beltrami operator $\Delta_n$ of the manifold $M^n$, with
eigenvalues $\mu_m^2$:
\begin{equation}\label{eigenvalueeq}
-\Delta_n\Psi_m=\mu_m^2\Psi_m
\end{equation}
The gravitational potential $V_{n+4}$ satisfies the Poisson
equation in $n+3$ spatial dimensions, when the Newtonian limit is
taken:
\begin{equation}\label{lap}
\Delta_{n+3}V_{n+4}=(n+1)\Omega_{n+2}G_{n+4}M\delta^{(n+3)}(x)
\end{equation}
with $M$, the mass of the system, $G_{n+4}$ the Newton constant in
$n+4$ dimensions and
\begin{equation}\label{omega}
\Omega_{n+2}=\frac{2\pi^{\frac{n+3}{2}}}{\Gamma(\frac{n+3}{2})}
\end{equation}
Equation (\ref{lap}) corresponds to the case of large compact
radius limit and has the solution:
\begin{equation}\label{neweq}
V_{n+4}=-\frac{G_{n+4}M}{r_n^{n+1}}
\end{equation}
In the case of our interest, where the compact dimensions have
small lengths, we find the harmonic expansion of $V_{n+4}$ in
terms of the eigenfunctions of the product space $M^4\times M^n$,
which reads:
\begin{equation}\label{harmonicexpansion}
V_{n+4}=\sum_m\Phi_m(r)\Psi_m(x)
\end{equation}
with $r$ denoting the coordinates of $M^4$ and $x$ denoting the
coordinates of $M^n$. Consequently, the $\Phi_m$ obey:
\begin{equation}\label{poiss3}
\Delta_3\Phi_m-\mu_m^2\Phi_m=(n+1)\Omega_{n+2}\Psi_m^*(0)G_{n+4}M\delta^{3}(x)
\end{equation}
with solution:
\begin{equation}\label{gravpot12}
\Phi_m(r)=-\frac{\Omega_nG_{n+4}M\Psi_m^*(0)}{2}\frac{e^{-|\mu_m|r}}{r}
\end{equation}
Finally the gravitational potential is written as:
\begin{equation}\label{fingravpot}
V_{n+4}=-\frac{\Omega_nG_{n+4}M}{2r}\sum_m\Psi_m^*(0)\Psi_m(x)e^{-|\mu_m|r}
\end{equation}
Since all point particles in the four dimensional spacetime have
no dependence on the internal compact space $M^n$, we can take
$x=0$ in (\ref{fingravpot}) to obtain the four dimensional
gravitational potential:
\begin{equation}\label{fingravpot1}
V_{4}=-\frac{G_{4}M}{r}\sum_m\Psi_m^*(0)\Psi_m(0)e^{-|\mu_m|r}
\end{equation}
which is valid for large values of $r$, compared to the lengths of
the compact dimensions.

In the above general result of relation (\ref{fingravpot1}) we
shall apply the eigenfunctions and eigenvalues of Sol manifolds.
For the eigenvalues we shall take the first two smallest
eigenvalues (and their corresponding eigenfunctions) since larger
eigenvalues are exponentially suppressed.

\section*{Sol Manifold modification of Newton's Law}
In order to compute corrections to the gravitational potential we
must solve equation (\ref{eigenvalueeq}) for the case of Sol
manifold $M_A^3$. A much more elaborate analysis of this section
can be found in \cite{bolsinov} We shall use the $(u,v,z)$
coordinates we introduced previously. The Laplace-Beltrami
operator for the manifold $M^3_A$ is:
\begin{equation}\label{Laplaceonsol}
\Delta=Ee^{2z\ln\lambda}\frac{\partial^2}{\partial
u^2}+2F\frac{\partial^2}{\partial u\partial
v}+Ge^{-2z\ln\lambda}\frac{\partial^2}{\partial
v^2}+\frac{\partial^2}{\partial z^2}
\end{equation}
which stems from the Riemannian metric (\ref{metricsol}). As usual
$E=|e_u|^2$, $F=|e_v|^2$ and $G=|(e_u,e_v)|$, where $e_u$ and
$e_v$ the basis of the $T^2$ lattice. Thus we must solve
\begin{equation}\label{laplacegener}
-\Delta \psi=E\psi
\end{equation}
Since the coefficients of $\Delta$ depend on $z$, the form of the
solutions of (\ref{laplacegener}) are of the form:
\begin{equation}\label{solutiongeneral}
\Psi_\gamma (u,v,z)=e^{2\pi i(\gamma,{\,}w)}f(z)
\end{equation}
where $\gamma$ is a vector of the dual lattice of $T^2$ and $w$ is
a vector of the $T^2$ lattice. Let us note that the vector of the
dual lattice $\gamma$ can be written
\begin{equation}\label{reciprocallatt}
\gamma=\frac{m}{R}e_u^*+\frac{n}{R}e_v^*
\end{equation}
with $e_u^*$ and $e_u^*$ the vector basis of the dual $T^2$
lattice. The vector of the $T^2$ lattice can be written:
\begin{equation}\label{ordlattice}
w=ue_u+ve_v
\end{equation}
The scalar product of $w$ and $\gamma$ is identified with the
usual product between the reciprocal and the Bravais lattice. Now
we shall use a proposition of reference \cite{bolsinov}, which
says that a function $\Psi=e^{2\pi i(\gamma,{\,}w)}f(z)$ satisfies
equation (\ref{laplacegener}) if and only if $f(z)$ satisfies the
modified Mathieu equation:
\begin{equation}\label{modmahtieu}
\Big{(}-\frac{d^2}{dz^2}+|\nu(\gamma)|\cosh2\mu(z+\alpha(\gamma))\Big{)}f(z)=\Lambda
f(z)
\end{equation}
with $\mu=\ln \lambda$, $\nu(\gamma)=8\pi^2CQ_{A^*}(\gamma)$ and
$\alpha(\gamma)=\frac{\ln\big{(}\sqrt{\frac{E}{G}}\frac{|(\gamma,e_u)|}{|(\gamma,e_v)|}\big{)}}{2\ln\lambda}$.
Also
\begin{equation}\label{c}
C=\frac{1}{\sqrt{D}\mathrm{Vol}(M_A^3)\sin\theta}
\end{equation}
The volume of the Sol manifold, ${Vol}(M_A^3)$, equals the volume
of the total bundle space $T^2\times S^1$. In the above $D$ is the
discriminant of the matrix $A$ defined in relation (\ref{discri1})
and $Q_{A^*}(\gamma)$ is the quadratic form (see \cite{bolsinov})
corresponding to the matrix $A^*$ acting to the dual lattice of
$T^2$, with
$Q_{A^*}(\gamma)=(\gamma,e_u)(\gamma,e_v)(\lambda-\lambda^{-1})$.
The eigenvalues $E$ and $\Lambda$ are related as follows:
\begin{equation}\label{eigenf}
E=\Lambda+\nu(\gamma)\cos\theta
\end{equation}
It is proved in reference \cite{bolsinov} that the functions
$\Psi_{\gamma}=e^{2\pi i(\gamma,{\,}w)}f_{\gamma}(z)$ form a
complete orthogonal basis on the Sol manifold. Also it is found
that the spectrum of the Laplace-Beltrami operator consists of two
parts:

\begin{itemize}
\item The trivial part, with eigenvalues
$E_k=\frac{4\pi^2k^2}{R^2}$, corresponding to $\gamma=0$ in the
dual lattice and to eigenfunctions $e^{i2\pi kz/R}$, $k=0,1,...$.
\item The non-trivial part with eigenvalues
$E_{k,[\gamma]}=\Lambda_k(\nu[\gamma])+\nu([\gamma])\cos\theta$
and eigenfunctions the solutions of (\ref{modmahtieu}).
\end{itemize}

In order to compute corrections to the gravitational potential due
to Sol structure of extra dimensions, we must compute the first
eigenvalues and eigenfunctions. We shall use the most interesting
case which is when $Vol(M_A^3)\sin\theta$ is large (or
equivalently $\nu\rightarrow 0$). Bear in mind that this term
contains the compactification radius of the extra dimensions (we
must state that the radii of all extra dimensions shall be
considered equal) and the deformation of the lattice in terms of
$\theta$. Following \cite{bolsinov} when $Vol(M_A^3)\sin\theta$
becomes large, $\nu(\gamma)$ becomes small. In that limit the
eigenvalues of the Laplace-Beltrami operator, stemming from the
non-trivial part, are $E_k=\Lambda_k$, with $\Lambda_k$ the
eigenvalues of the modified Mathieu operator,
\begin{equation}\label{modopoe}
M=-\frac{d^2}{dz^2}+|\nu(\gamma)|\cosh2\mu z
\end{equation}
We order the first eigenvalues as $E_0=0\leq E_1\leq E_2\leq...$.
We shall use the first two, since the corrections to the
gravitational potential fall exponentially at the eigenvalues grow
larger. Thus the first two are $E_0=0$ and $E_1$, which
asymptotically reads:
\begin{equation}\label{firsteigenvalue}
E_1\sim \frac{(\ln\lambda)^2\pi^2}{(\ln C)^2}
\end{equation}
with $C$ as before:
\begin{equation}\label{c1}
C=\frac{1}{\sqrt{D}\mathrm{Vol}(M_A^3)\sin\theta}
\end{equation}
The eigenfunctions corresponding to the Mathieu operator
(\ref{modopoe}) are in general,
\begin{equation}\label{momathfunc}
f_{m}(z)=\sum_{k=0}^{\infty}A_{2k}^{2m}\cosh[2kz]
\end{equation}
with $'m'$ counting the eigenvalues, $m=0$ corresponds to $E_0$
e.t.c.

Now we are ready to compute the modifications to the gravitational
potential coming from the first two eigenvalues of the Laplace
operator on Sol manifolds. Thus we substitute $E_0$ and $E_1$ in
relation (\ref{fingravpot1}) with $|\mu_0|=\sqrt{E_0}$ and
$|\mu_1|=\sqrt{E_1}$. Also we substitute the eigenfunctions for
the Sol manifold, thus:
\begin{equation}\label{eigenzero}
 \Psi_m(0)=\sum_{k=0}^{\infty}A_{2k}^{2m}
\end{equation}
The asymptotic behavior of the coefficients $A_{2k}^{2m}$ for
$\nu\rightarrow 0$ (which is the most interesting case and we
shall dwell on it) is really simple \cite{mclachan}. It turns out
that the terms of the form $A_m^{(m)}$ tend to $1$, while terms of
the form $A_m^{(k)}$, with $m\neq k$ tend to zero. Thus the
gravitational potential of relation (\ref{fingravpot1}) becomes:
\begin{equation}\label{fingravpot13}
V_{4}=-\frac{G_{4}M}{r}\sum_m\sum_{k=0}^{\infty}A_{2k}^{2m}A_{2k}^{2m}e^{-|\mu_m|r}
\end{equation}
So for the first two eigenvalues we have
\begin{equation}\label{lastexpr}
V_{4}=-\frac{G_{4}M}{r}\Big{(}A_{0}^{0}A_{0}^{0}e^{-|E_0|r}+A_{2}^{2}A_{2}^{2}e^{-|E_1|r}\Big{)}
\end{equation}
or (using the asymptotic behavior for the coefficients
$A_m^{(m)}$):
\begin{equation}\label{lastexpr1}
V_{4}=-\frac{G_{4}M}{r}\Big{(}e^{-\sqrt{E_0}r}+e^{-\sqrt{E_1}r}\Big{)}
\end{equation}
Since $E_0=0$ and using relations (\ref{firsteigenvalue}) and
(\ref{c1}) we obtain finally:
\begin{equation}\label{lastexpr2}
V_{4}\sim
-\frac{G_{4}M}{r}\Big{(}1+e^{-\big{|}\frac{(\ln\lambda)\pi}{(\ln
C)}\big{|}r}\Big{)}
\end{equation}
In the relation above it is clear that the Sol manifold correction
to the Newton's law gravitational potential depends on the
compactification radii of the extra dimensions $R$, the angle
$\theta$ of the vectors $e_u$ and $e_v$ and on the eigenvalues of
the hyperbolic gluing map $A$. In the next section we shall
investigate the parameter space of the corrections found (which is
very rich) and we shall compare it with the $T^3$ manifold
corrections, to see how much different these are.

\section*{Analysis of the parameter space and comparison with the
$T^3$ corrections}

\begin{figure}[h]
\begin{center}
\includegraphics[scale=.7]{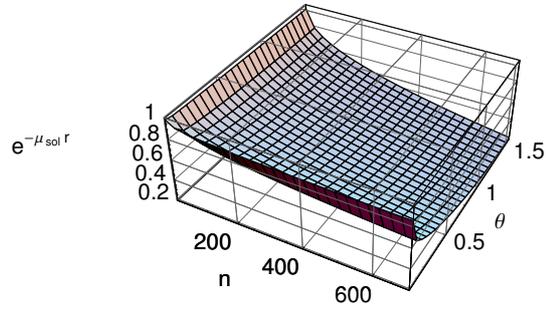}
\end{center}
\caption{Dependence of the range $e^{-\mu_mr}$ of Sol corrections
on $\theta$ and $n$ with $r=0.1{\,}mm$, $R=0.05{\,}mm$}
\label{3d1}
\end{figure}

\begin{figure}[h]
\begin{center}
\includegraphics[scale=.7]{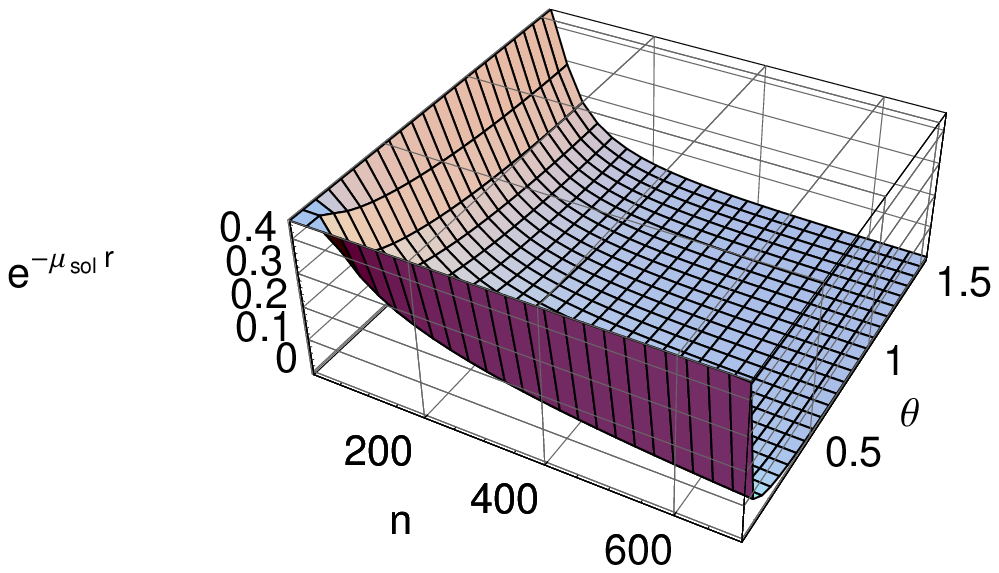}
\end{center}
\caption{Dependence of the range $e^{-\mu_mr}$ of Sol corrections
on $\theta$ and $n$ with $r=0.1{\,}mm$, $R=0.1{\,}mm$}\label{3d2}
\end{figure}

\begin{figure}[h]
\begin{center}
\includegraphics[scale=.8]{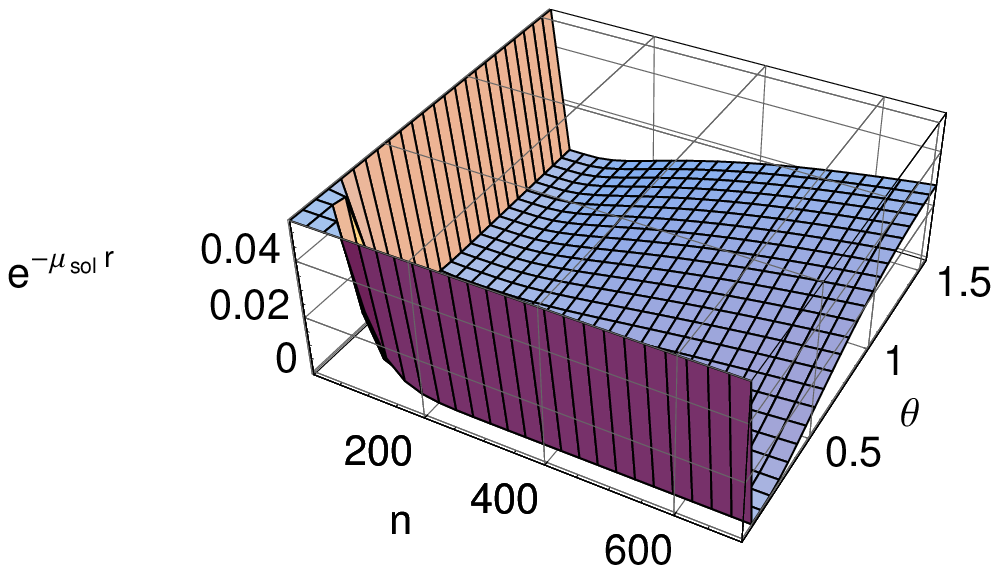}
\end{center}
\caption{Dependence of the range $e^{-\mu_mr}$ of Sol corrections
on $\theta$ and $n$ with $r=0.1{\,}mm$, $R=0.2{\,}mm$}\label{0201}
\end{figure}

The parameter space of the Sol correction to the gravitational
potential is very rich. Let us examine first the dependence of the
Sol correction range $e^{-\mu_mr}$ on the parameters $\theta$ and
$n$. In Figure (\ref{3d1}) we plot the dependence for the
compactification radius value $R=0.05{\,}mm$ (smaller than the
current experimental bound $r=0.2mm$) and for $r=0.1{\,}mm$, while
in Figure (\ref{3d2}) and (\ref{0201}) the values of $R$ are
$0.1{\,}mm$ and $0.2{\,}mm$ respectively. We tried to check the
plots near the experimental bound for extra dimensions. It seems
that the Sol structure gives very large corrections to gravity if
the compactification radius is very small (of order $R\sim
10^{-6}{\,}$ and smaller). We shall study the interesting cases
which have compactification radius around $R\sim 0.2{\,}mm$. Also
we shall check out the behaviour of the range of the corrections
around $R\sim 10^{-8}m$ which is the expected scale that three
extra dimensional spaces should have. Looking in Figures
(\ref{3d1}), (\ref{3d2}), and (\ref{0201}), we can see that the
last two have similar dependence. It seems that for large values
of $n$ ($>200$) and for $\theta>0.3$, the corrections are very
small. In the other two graphs when $n>300$ and for small values
of $\theta$, the corrections are small. In the following we shall
investigate some interesting cases stemming from the above three.
To have a clear picture we shall compare the range of Sol
manifolds with the range of the $T^3$ torus corrections.

\begin{figure}[h]
\begin{center}
\includegraphics[scale=.7]{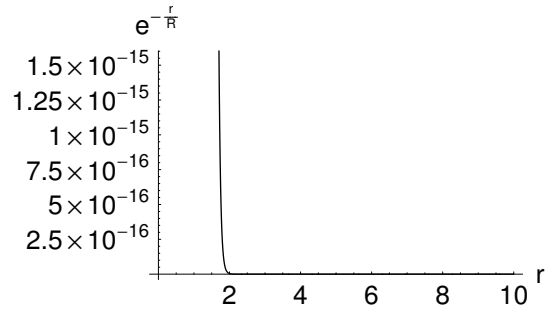}
\end{center}
\caption{$r$-dependence of the range of $T^3$ corrections for
$R=0.05{\,}mm$}\label{t3}
\end{figure}

\begin{figure}[h]
\begin{center}
\includegraphics[scale=.7]{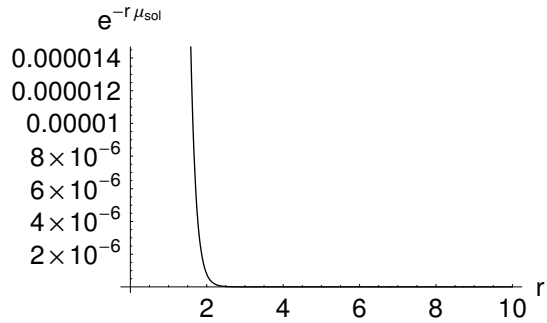}
\end{center}
\caption{$r$-dependence of the range $e^{-\mu_mr}$ of Sol
corrections with $R=0.05{\,}mm$, $n=250$ and
$\theta=\frac{\pi}{3}$}\label{005}
\end{figure}

 In Figure (\ref{t3}) we plot the range of the $T^3$ corrections as a function of $r$ with $R=0.05mm$ and
$\theta=\pi/3$ and in Figure (\ref{005}) the corresponding
dependence for the Sol manifold case. As expected from Figure
(\ref{3d1}), the term $e^{-\mu_mr}$ is very big compared to
$e^{-r/R}$, for small $n$ ($\sim 5$). As $n$ grows the term
becomes smaller and smaller. Figure (\ref{005}) corresponds to
$n=250$. As $n$ grows, the range of Sol-corrections becomes
comparable and after a value of $n$, smaller compared to the $T^3$
corrections. The small $\theta$ case is similar. Concluding the
above investigation we could say that the Sol range values are big
for the above case in which the corresponding torus range values
were small.

The case with $R=0.2${\,}mm is very interesting. As we can see in
Figure (\ref{comparissonnew}) the range of Sol corrections can
vary significantly depending on the value that $n$ takes. Compared
to the corresponding 3-Torus range, Sol corrections can be much
larger (for small $n$) and comparable (for $n>450$) but never
smaller even for very large $n$.

Another result is that general Sol corrections can exist for
compactification radii and at distances for which the
corresponding $T^3$ torus range values are very large (and
consequently not experimentally preferable).

\section*{Discussion}
In extra dimensional models, gravity behaves exactly as in four
dimensional models when long distances are considered. When
however we are close at the scale of the extra dimensions, the
potential changes and we expect macroscopical changes in the
strength and range of the gravitational force. This is expressed
by the relation
\begin{equation}\label{massscale}
M_{Pl}^2=M_*^{n+2}(2\pi R)^n
\end{equation}
Depending on the topology and geometry of the extra dimensions,
the corrections may vary.

There exist several bounds constraining the size and number of
extra dimensions. For gravitational corrections the strongest
constraint comes from the E\"{o}t-Wash experiment. The results
show consistency with Newtonian gravity down to 200 microns
\cite{hoyle}.

In the ADD model the scale $M_*$ is taken to be 1 TeV. For this
value we obtain the following sizes for extra dimensions with the
same radius $R$:
\begin{center}
\begin{tabular}{|c|c|}
  \hline
  number of extra dimensions & R (m) \\
  \hline
  n=1 & $\sim 10^{12}$ \\
  \hline
  n=2 & $\sim 10^{-3}$ \\
  \hline
  n=3 & $\sim 10^{-8}$ \\
\hline
\end{tabular}
\end{center}
\begin{figure}[h]
\begin{center}
\includegraphics[scale=.8]{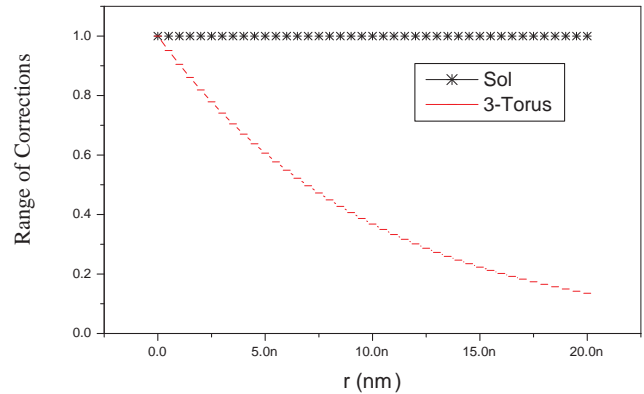}
\end{center}
\caption{$r$-dependence of the range of $T^3$ and Sol corrections
for $R=10{\,}nm$} \label{comparisson}
\end{figure}
The case of one extra dimension with quantum gravity scale $\sim$1
TeV is ruled out by solar system experiments. Also two extra
dimensions are ruled out since for $M_*$=1 TeV the experiments
have tested Newton's force down to 200 microns.

\noindent The ADD $n=3$ case is not excluded yet with $M_*=1$ TeV.

Bounds on extra dimensions are posed when gravitons are taken into
account. Gravitons would be copiously produced in colliders unless
$M_*$=1 TeV for $n=3$. There exist stronger bounds when light
gravitons are taken into account. These are emitted from
supernovae and the astrophysical bounds for three extra dimensions
on $M_*$=1 TeV are of the order TeV \cite{cullen}.

\begin{figure}[h]
\begin{center}
\includegraphics[scale=.8]{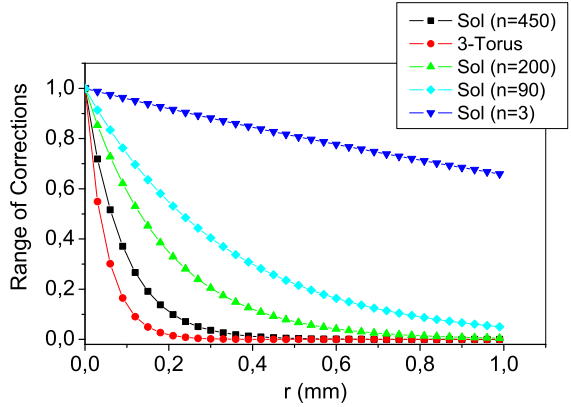}
\end{center}
\caption{Comparison of the $r$-dependence of the range of $T^3$
and Sol corrections for $R=0.2{\,}${\,}mm for various $n$, and
$\theta=\pi/3$} \label{comparissonnew}
\end{figure}

\begin{figure}[h]
\begin{center}
\includegraphics[scale=.8]{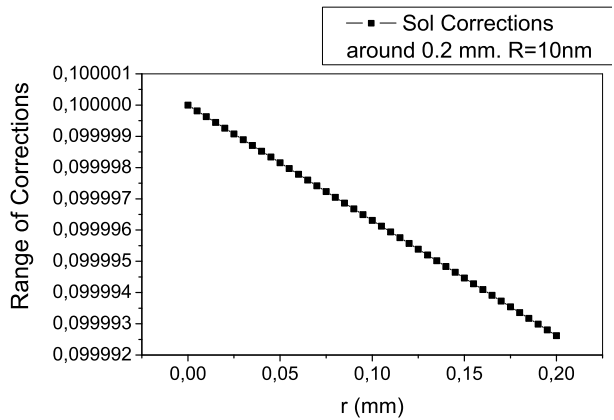}
\end{center}
\caption{$r$-dependence of the range of Sol corrections for
$R=10{\,}nm$ and $r\sim 10^{-4}$m} \label{mikromegalo1}
\end{figure}

In the case of Sol manifolds the same arguments hold and also, due
to the toroidal fibration, equation (\ref{massscale}) is valid
(the only difference might be in some graviton cross-sections
because the distribution of eigenvalues of the Laplacian on Sol
manifolds is such that leads to non-constant mass splitting. In
fact the eigenvalue distribution is not Poisson, as for most toric
topologies happens. This will be studied elsewhere). Sol space is
three dimensional. Taking into account that the experimental and
theoretical bound is $R=10${\,}nm, we compare the range of the
three torus corrections and the range of Sol manifolds
corrections. In Figure (\ref{comparisson}) we plot the range of
the two manifolds.
\begin{figure}[h]
\begin{center}
\includegraphics[scale=.9]{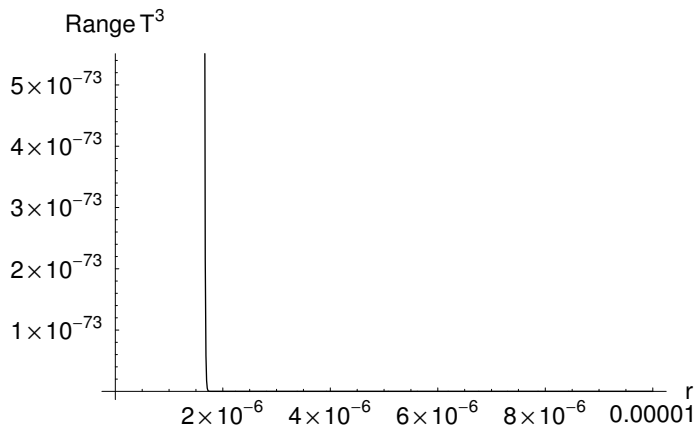}
\end{center}
\caption{$r$-dependence of the range of $T^3$ corrections for
$R=10{\,}nm$ and $r\sim 10^{-6}$m} \label{aps}
\end{figure}
As we can see, the range of the Sol corrections is significantly
larger from the three torus corrections. Also the range of the
$T^3$ corrections drop much faster compared to the Sol case. This
is very interesting phenomenologically. In the Sol case, the
values of the corrections are too close one to each other and in
Figure (\ref{comparisson}) appears as a straight line (but it is
not). In addition we found that the values of $n$ (the parameter
appearing in the $SL(2,Z)$ matrix) and of $\theta$, do not seem to
modify the results significantly, for $R\sim 10^{-8}{\,}${\,}m.
The last is to be contrasted with the case $R\sim 0.2${\,}mm as
can be seen in Figure (\ref{comparissonnew}).
As we can see the $n$ value modifies significantly the range of
Sol corrections when $R\sim 0.2${\,}mm. For large $n{\,}(\sim
450)$ the Sol corrections are comparable to the 3-torus
corrections while for small $n$ the range of Sol corrections grow
larger and larger.
\begin{figure}[h]
\begin{center}
\includegraphics[scale=.7]{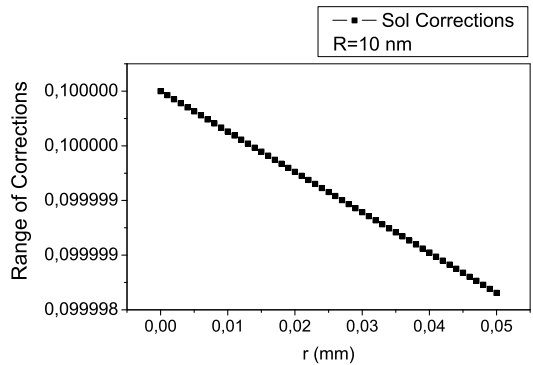}
\end{center}
\caption{$r$-dependence of the range of Sol corrections for
$R=10{\,}nm$ and $r\sim 10^{-6}$m} \label{mikromegalo2}
\end{figure}
Let us return to the case $R\sim 10^{-8}${\,}m and see how the
modifications behave down to the $200{\,}\mu$m{\,}($0.2${\,}mm).
We found that in the 3-torus case, the range of the corrections
becomes extremely small for distances $r\sim 10^{-6}$m and
smaller, while the Sol corrections for $r\sim 10^{-6}$m still
remain significantly larger compared to the torus case, as can be
seen in Figures (\ref{aps}) and (\ref{mikromegalo2}). In the case
of Sol corrections we can see that around $10^{-4}${\,}m and
lower, the corrections become small as can be seen in
Figure~\nobreak(\ref{mikromegalo1}) (although the range still
remains significantly larger compared to the 3-Torus case).

A more complete investigation for small $R$ would require a
numerical calculation of the eigenvalues of the Sol manifold
Laplace-Beltrami operator. There exists an algorithm that
calculates numerically the eigenvalues of three dimensional
manifolds and can be found in \cite{numerically} and references
therein.

\newpage

\section*{Conclusions}
In this letter we studied the corrections to the gravitational
potential caused by an Sol manifold extra dimensional compact
space. Specifically we used the first two eigenvalues of the
Laplace-Beltrami operator on the Sol manifold which showed us that
the Yukawa type modifications are affected by the parameters $n$,
$\theta$ and $R$, where $R$ the compactification radius and
$\theta$ the angle between the basis vectors of the $T^2$ lattice.
The parameter $n$ is connected with the matrix $A$ $\epsilon$
$SL(2,Z)$, which is hyperbolic and defines the gluing of the torus
fibers. After investigating the parameter space we found
interesting results, mainly concluding that the range of Sol
structure corrections can be similar to the $T^3$ torus results
and (more interestingly) can be very different to the $T^3$
results depending to the values of the parameters. This is
valuable phenomenologically as can be seen in the previous
sections.

Also let us discuss the fact that for Sol manifolds, when $R$ is
small, the range of the corrections is large compared with the
3-torus case. This must have to do with the intrinsic
hyperbolicity of Sol manifolds. This intrinsic hyperbolicity has
effect on the eigenvalues and their distribution. Due to this if
we wish to take the limit $R\rightarrow \infty$ we must be very
careful. In the limit $R\rightarrow \infty$ the eigenvalues grow
smaller and smaller much more faster than in the 3-torus case.
Thus roughly we could say that Newton's law is recovered. However
for large $R$ we must take into account the eigenvalues we
disregarded since we still work in a compact space but $R$ is not
small anymore. Some eigenvalues of Sol (not trivial to be found)
can be found in Ref.\cite{bolsinov}.

As a final remark let us note that the values of the range of the
Sol corrections remain very large compared to the 3-Torus ones
(almost flat even at $R\sim 0.2$mm !). This happens even near the
experimental limits $0.2${\,}mm. At first this may appear very
strange but the study involved only the range of the corrections,
so the large range values might be interesting feature since the
range is divided by the distance $r$. Thus in some cases the
corrections might be measurable. However extra care must be taken
when $R$ is small, as we mentioned previously.

Finally we left unanswered two issues, the graviton production and
how do we distinguish Sol manifolds corrections from other
corrections (the last due to the rich parameter space of Sol
manifolds). We hope we address these in the future.

\section*{Acknowledgements}
V.O would like to thank Alex Kehagias for reading the manuscript
and for valuable comments and prospective extensions of the above
he gave. Also would like to thank both the referees for invaluable
comments and suggestions that improved significantly the quality
of the paper.

\newpage

\newpage

\end{document}